\documentclass{PoS}
\def\LP{\left(}         % left parenthesis
\def\RP{\right)}        % right parenthesis
\newcommand{\BE}{\begin{displaymath}}
\def\EE{\end{displaymath}}
\def\BNE{\begin{equation}}
\def\ENE{\end{equation}}
\def\BEA{\begin{eqnarray}}
\def\EEA{\nonumber\end{eqnarray}}

\title{Baryon masses with improved staggered quarks}

\ShortTitle{Baryon masses with $a_{tad}^2$ staggered quarks}

\author{C.~Bernard$^a$, C.~Davies$^b$, C.~DeTar$^c$, Steven Gottlieb$^d$, U.M.~Heller$^e$,
   J.E.~Hetrick$^f$, L.~Levkova$^c$, J.~Osborn$^g$, D.B.~Renner$^h$, R.~Sugar$^i$
   and \speaker{D.~Toussaint}$^h$\\
        \llap{$a$} Physics Department, Washington University, St. Louis, MO 63130, USA\\
        \llap{$b$} Physics Department, Glasgow University, Glasgow, G12 8QQ, UK\\
        \llap{$c$} Physics Department, University of Utah, Salt Lake City, UT 84112, USA\\
        \llap{$d$} Physics Department, Indiana University, Bloomington, IN 47405, USA\\
        \llap{$e$} American Physical Society, One Research Road, Ridge, NY 11961-9000, USA\\
        \llap{$f$} Physics Department, Pacific University, Stockton, CA 95211, USA\\
        \llap{$g$} Physics Department, Boston University, Boston, MA 02215, USA\\
        \llap{$h$} Physics Department, University of Arizona, Tucson, AZ 85721, USA\\
        \llap{$i$} Physics Department, University of California, Santa Barbara, CA 93106,
USA\\
        E-mail:
\email{cb@wuphys.wustl.edu},
\email{c.davies@physics.gla.ac.uk},
\email{detar@physics.utah.edu},
\email{sg@indiana.edu},
\email{heller@aps.org},
\email{jhetrick@uop.edu},
\email{ludmila@physics.utah.edu},
\email{josborn@physics.bu.edu},
\email{dru@physics.arizona.edu},
\email{sugar@physics.ucsb.edu},
\email{doug@physics.arizona.edu}
	}

\abstract{
The MILC collaboration's simulations with improved staggered quarks are
being extended with runs at a lattice spacing of 0.06 fm with quark masses
down to one tenth the strange quark mass.  We give a
brief introduction to these new simulations and the determination of the
lattice spacing.  Then we combine these new runs with older results to
study the masses of the nucleon and the $\Omega^-$ in the continuum and
chiral limits.
          }

\FullConference{The XXV International Symposium on Lattice Field Theory\\
		 July 30 - August 4 2007\\
		 Regensburg, Germany}

\begin{document}

\section{Introduction}

In this work we update our determination of the lattice spacing in the
MILC collaboration's program of QCD simulations using $N_f=2+1$ flavors
of dynamical staggered quarks with the $a_{tad}^2$ action\cite{ASQTAD},
and update our calculations of the nucleon and $\Omega^-$ masses.
Determining the lattice spacing is central to any simulation program.
We emphasize that the same lattice spacing determination, with the same
uncertainties, affects everything computed from these lattices.
The spectrum of the light quark hadrons is an important test of a lattice
QCD simulation, and is a standard part of all large lattice QCD
programs.
Getting these well known masses right gives us confidence
in our ability to predict unknown masses and matrix elements.
Both the nucleon and $\Omega^-$ are stable against strong decays and
well known experimentally, and hence are good tests of our program.
Our previously published work on the light quark spectrum and lattice
spacing determination is in Refs.~\cite{MILCSPECTRUM1,MILCSPECTRUM2}.

We are now doing simulations with a smaller lattice spacing, $a\approx 0.06$ fm,
in addition to lengthening some runs and adding runs at other quark masses
at larger lattice spacing.   The runs at $a\approx 0.06$ fm are all still in
progress, so all results reported here are preliminary.
Table I shows the simulation parameters for the $a\approx 0.06$ fm runs.

\begin{table}[h!] \begin{center}
\setlength{\tabcolsep}{3mm}
\begin{tabular}{|c|c|c|c|r|c|c|}
\hline
%\noalign{\vspace{0.15cm}}
$am_q$ / $am_s$  & \hspace{-1.0mm}$10/g^2$ & size & volume & number & $a$ (fm) & Algorithm \\
\hline
%\noalign{\vspace{0.15cm}}
0.0072  / 0.018   & 7.48 & $48^3\times144$ & $(2.9\;{\rm fm})^3$ & 555 & 0.060 & R \\
0.0036  / 0.018   & 7.47 & $48^3\times144$ & $(2.9\;{\rm fm})^3$ & 480 & 0.060 & RHMD-1 \\
0.0018  / 0.018   & 7.46 & $64^3\times144$ & $(3.8\;{\rm fm})^3$ & 46 & 0.061 & RHMD-4 \\
\hline
\end{tabular}
\end{center}
\caption{\label{tab:lattices}
%\vspace{1cm}
Lattice parameters.
Lattice spacings come from the smoothed $r_1/a$ values and $r_1=0.318\;$fm.
``RHMD-$n$'' is a RHMD algorithm using $n$ pseudofermion fields.  Details
of the algorithms will be given elsewhere.
}
\end{table}

\section{Finding the lattice spacing}

The results of a lattice computation are in units of the lattice spacing
$a$, so a determination of $a$ in physical units is essential.  This requires
taking some physical quantity (or average over quantities) as a standard.
At this time we are using the $\Upsilon$ mass splittings determined by
the HPQCD group\cite{HPQCD_UPSILON}.
A possible alternative is $f_\pi$, and we are investigating whether this
would be better\cite{BERNARD_LAT07}.
Because $\Upsilon$ mass splittings are not available on all of our ensembles
and because the static quark potential can be accurately determined,
we use $r_1$  defined by $r_1^2 F(r_1) = 1.0$ \cite{SOMMER,MILC_POTENTIAL}
as a quantity to interpolate the lattice spacing, as described in Ref.~\cite{MILCSPECTRUM2}.
Thus, the $\Upsilon$ splittings give $r_1=0.318(7)$ fm , which is then used to
determine the lattice spacing for all gauge couplings and quark masses.

Static quark potentials were calculated from correlators of time direction lines,
with Coulomb gauge fixing for the spatial connections\cite{MILC_POTENTIAL}.
As the lattice spacing decreases, determining the static potential becomes
more difficult because of the increasing constant term, corresponding
to the singular self-energy of the static quark.
Since the Wilson loop expectation value falls off exponentially
as $\langle W(R,T)\rangle \approx \exp\LP -V(R)T \RP$, a larger
constant in $V(R)$ translates into an exponentially smaller
expectation value of $W(R,T)$, and larger fractional statistical
error.
This can be alleviated,
at the cost of some spatial resolution, by smearing the time direction
links in the lattice.

Figure~1a shows the static quark potential, with and
without one iteration of APE smearing of the time direction links,
while Fig.~1b shows the static quark
potential at $m_q = 0.4 m_s$ for four different lattice
spacings.

\begin{figure}[h]
\label{fig.pot1}
\begin{center}
\parbox[t]{15cm}{
\parbox[t]{7.0cm}{
\includegraphics[width=7.0cm]{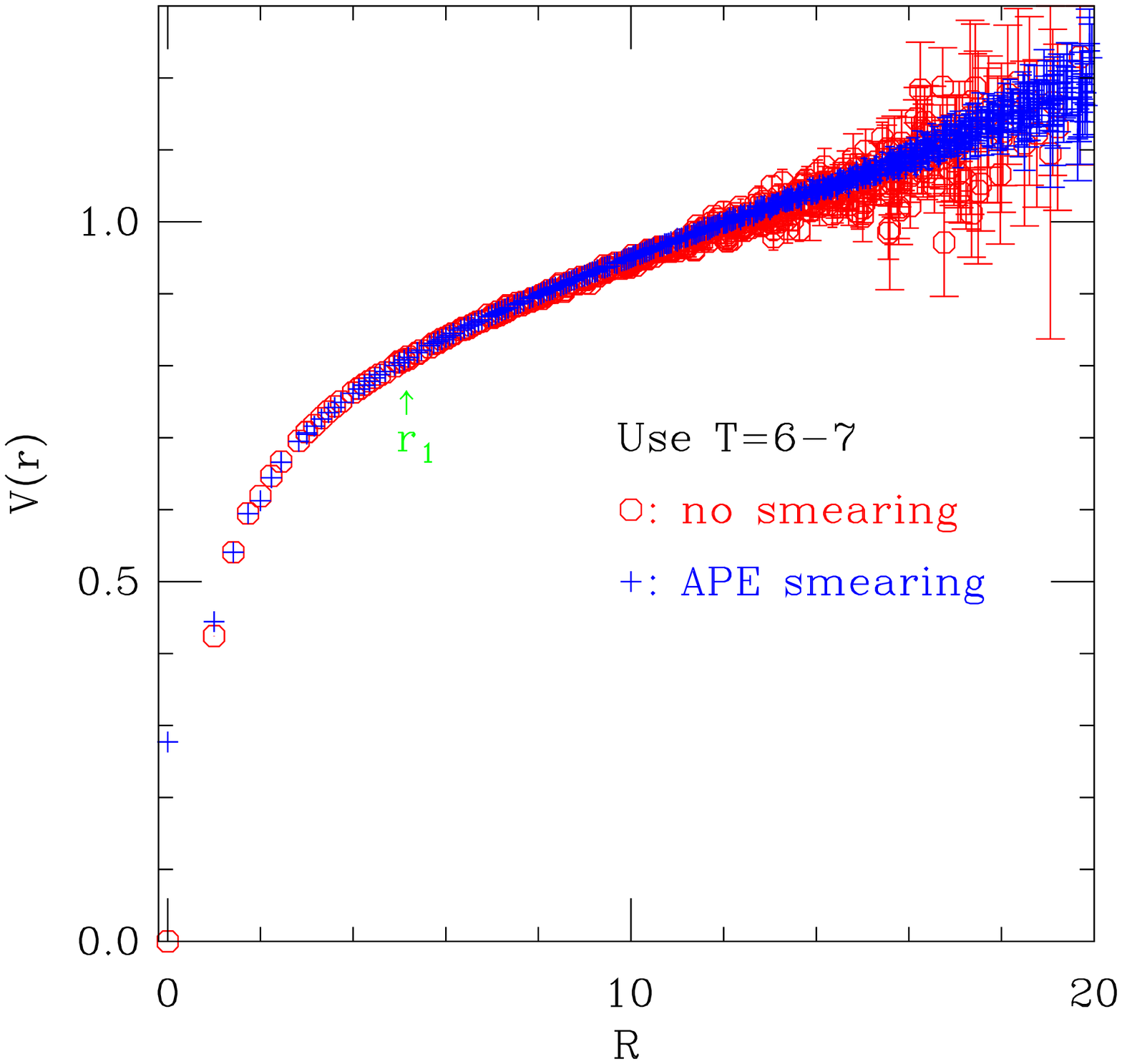}
Figure 1a:
The static quark potential, with and
without APE smearing of the
time direction links.
This is for the $am_q/am_s=0.0072/0.018$ run.
The constants in the potential differ (that's the point), so we
added 0.27731 to the smeared potential to show the agreement.
(This shift forces the potentials to agree at $r_1$.)
}
\hspace{1.0cm}
\parbox[t]{7.0cm}{
\includegraphics[width=7.0cm]{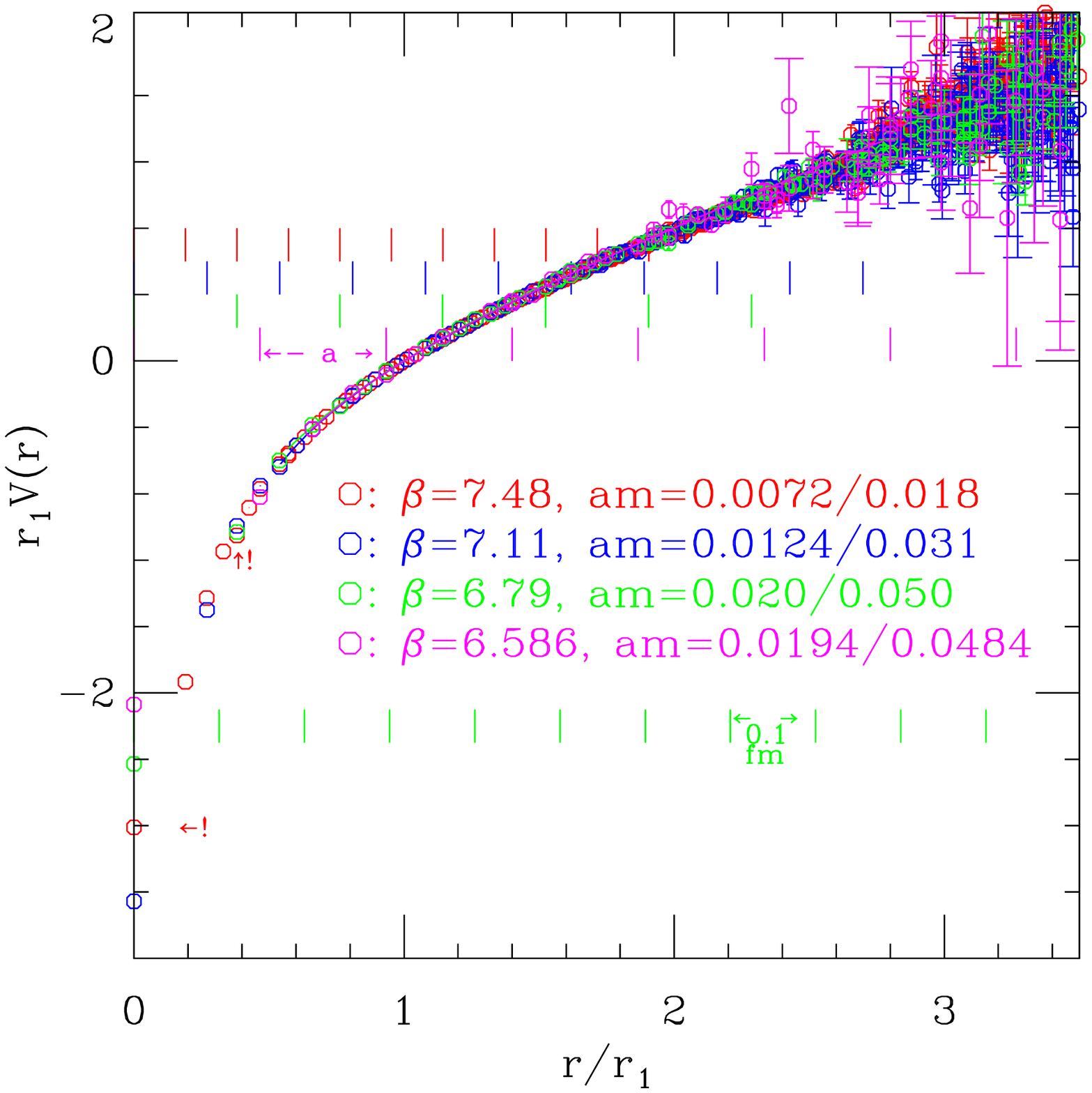}
Figure 1b:
The static quark
potential at $m_q = 0.4 m_s$ for four different lattice
spacings, roughly $a=0.06$, $0.09$, $0.125$ and $0.15$ fm.
The $a=0.06$ fm potential uses smearing of the time links.
The colored rulers near $V=0$ are in units of the corresponding
lattice spacings, while the ``physical'' ruler near $V=\,-2$ is
in units of $0.1$ fm.
}
}
\end{center}
\end{figure}

In Fig.~\ref{fig.pot1}b note the visible lattice artifact at the
on-axis $(2,0,0)$ point marked
with $\uparrow !$.
The spatial range used for the $r_1$ determination at $a\approx 0.06$ fm was
4.01 to 7.01.
Also note that the $a=0.06$ fm potential (red)
at $r=0$ (marked with $\leftarrow !$) is much
smaller in magnitude than if it continued the trend, again showing the
effect of the smearing.

\begin{figure}[h]
\begin{center}
\parbox[t]{15cm}{
\label{fig.test}
\parbox[t]{7.0cm}{
\includegraphics[width=7.0cm]{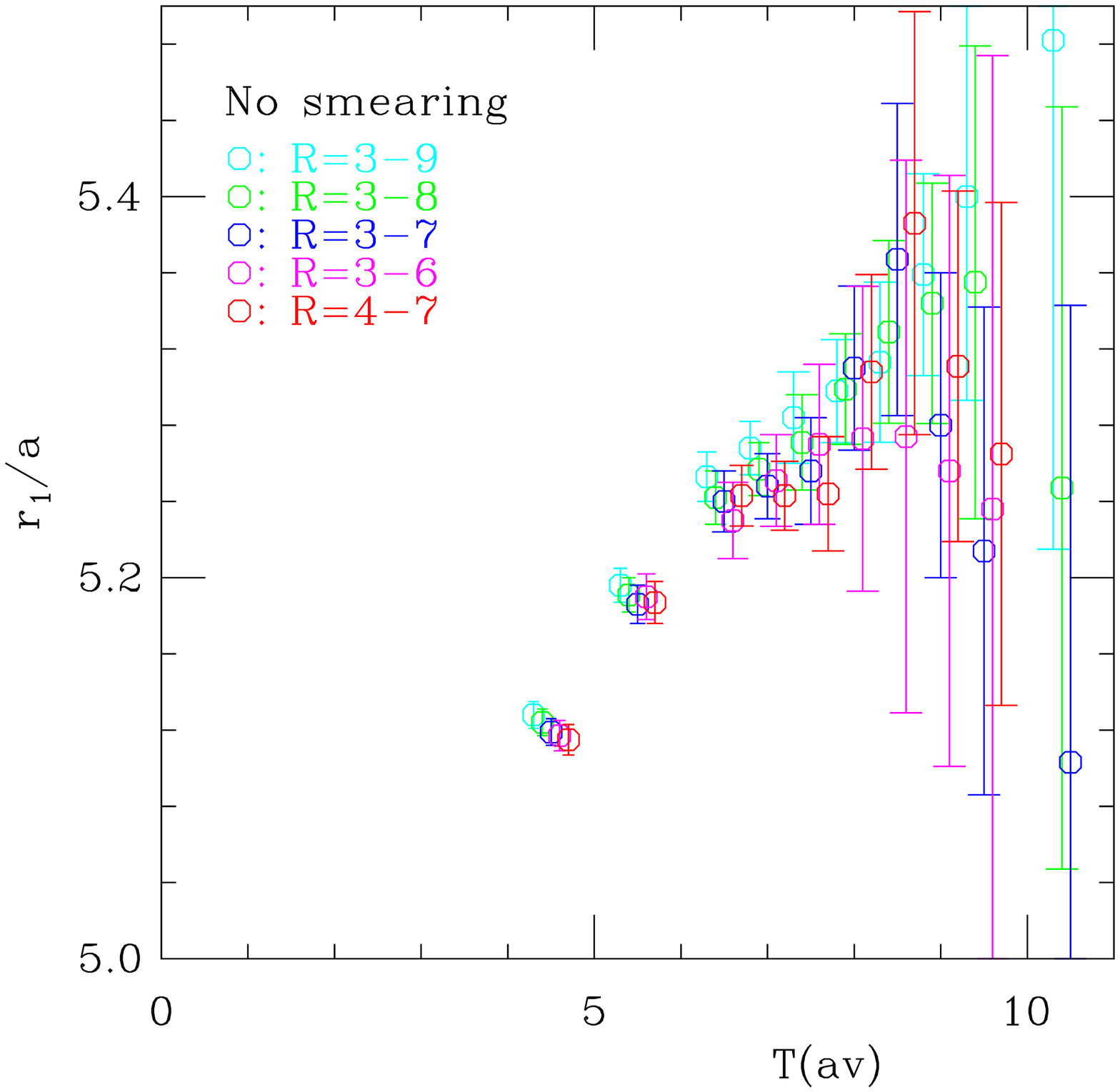}
Figure 2a:
Estimates of $r_1/a$ from ``loops'' with unsmeared time links.
This is for the $am_q/am_s=0.0072/0.018$ run.
}
\hspace{1.0cm}
\parbox[t]{7.0cm}{
\includegraphics[width=7.0cm]{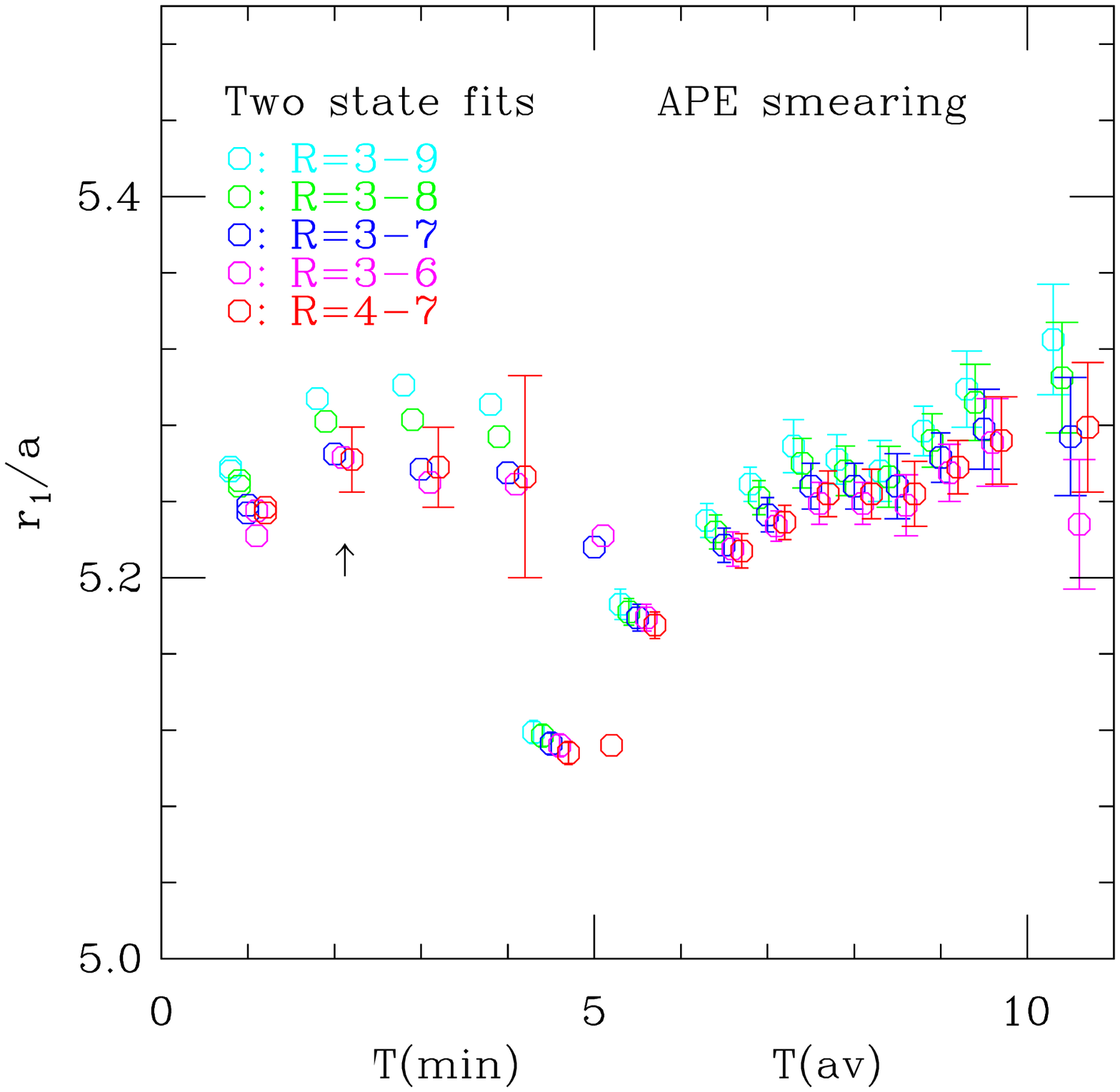}
Figure 2b:
Estimates of $r_1/a$ from the same run, with one iteration of
APE smearing including projection onto SU(3) for the time
direction links (right side).
Fits including an excited state (left side).
The arrow indicates the point taken as our estimate.
}
}
\end{center}
\label{fig.pot2}
\end{figure}

To find $r_1$ the potential was fit to $V(r) = C + A/r + B r$ for a range around $r_1$.
With $a\approx0.06$ fm, lattice artifacts are unimportant in these distance ranges.
Errors are from a jackknife analysis, using a block size of 20 lattices, or
100 molecular dynamics time units.
Figure~2a shows estimates of
$r_1/a$ from various spatial and temporal distance ranges without
smearing, while Fig.~2b (right half) shows fits with one iteration of APE smearing,
including projection onto SU(3).
The fits shown in Figs.~2a and 2b (right) are from fitting the ``Wilson loops''
to a single exponential: $V(r) = \log( W(r,T)/W(r,T+1) )$.
We also do fits including an excited state of the potential.  This gives
less dependence on fit range (smaller systematic error) but larger
statistical errors.  The $r_1/a$ estimates on the left side of Fig.~2b
are from such two-state fits.
%
%\begin{center}
%\parbox[t]{15cm}{
%\parbox[t]{7.0cm}{
%\includegraphics[width=7.0cm]{r1_748_comb.ps}
%Fits including an excited state (left side), compared to the one state
%fits at larger distance (right side).
%}
%\hspace{1.0cm}
%\parbox[t]{7.0cm}{
%\includegraphics[width=7.0cm]{r1_747_comb.ps}
%The same, except this is for the $am_q=0.0036/0.018$ run.
%}
%}
%\end{center}
%
Fitting this, together with our previous results at $a\approx 0.018$,
$0.015$, $0.0125$ and $0.09$ fm, we get a ``smoothed $r_1$'',
which is then used to set the length scale for all ensembles.

%Fitting this, together with our previous results at $a\approx 0.018$,
%$0.015$, $0.0125$ and $0.09$ fm, we get a ``smoothed $r_1$''.
%\BNE
%\log(r1/a) = C_{00} +
%C_{10} (\beta-7.0)+C_{20} (\beta-7.0)^2+C_{30} (\beta-7.0)^3 +
% C_{01} M + C_{02} M^2
%\ENE
%\BEA
%C_{00} &=& 1.2731(40) \ \ \ 
%C_{10} = 0.9346(65) \ \ \ 
%C_{01} = -1.141(92) \EL
%C_{20} &=& -0.171(10) \ \ \ 
%C_{02} = 1.20(46) \ \ \ 
%C_{30} = 0.152(31) \EL
%\chi^2/D &=& 43.1/32
%\EEA
%(The parameters are all correlated.)
%Here $M=2 am_l + am_s$ is the sum of the quark masses.

\section{Nucleon and $\Omega^-$ masses}

Both the nucleon and the $\Omega^-$ are stable against strong decays, and so
their masses should be accessible in a straightforward way to lattice calculations.
In principle the nucleon mass could be used to determine the lattice spacing.
In practice its correlators are noisy, and the non-trivial chiral
extrapolation
makes an accurate determination difficult.
The $\Omega^-$ has a very mild chiral extrapolation --- no terms with
$\log(m_\pi)$
or $m_\pi^3$ at one loop order.
Since it contains three valence strange quarks, its mass is very sensitive to
the strange quark mass.
Therefore, the $\Omega^-$ mass checks our determination
of the lattice strange quark mass from pseudoscalar physics.

Figure~3a shows the nucleon correlators from the $a\approx 0.06$ fm runs at $0.4 m_s$
and $0.2 m_s$.
Note the expected difference in the slope, the alternating (opposite
parity) signal at short distances, and the increasing error bars at large distances.
Figure ~3b shows fits to one of these correlators ($m_q \approx 0.2 m_s$) as well as correlators
for roughly the same light quark mass at $a\approx 0.125$ and $0.09$ fm.
In this
plot the size of the symbol is proportional to the confidence level of the fit, with
the symbol size used in the captions corresponding to 50\%.   From graphs like
this we look for a plateau and reasonable confidence level ($\chi^2$).  The arrows
in the graph indicate the fits that were chosen.  Most of the difference in the
masses comes from the fact that the light quark masses in these three runs were
not exactly equivalent, and some comes from effects of lattice discreteness.

In Figure~4a we show nucleon masses for $a\approx 0.0125$ fm, $0.09$ fm and $0.06$ fm,
where the three colored arrows in Fig.~3b indicdate the fits that were
chosen to include in this figure.
The black error bar at the left is the experimental value.  The error
on this comes from the uncertainty in $r_1$ (We use $0.318(7)$ fm).
Another point at $a\approx 0.06$ fm at $m_l \approx 0.1 m_s$ ($(m_\pi r_1)^2 \approx 0.12$)
is in progress.

\begin{center}\parbox[t]{15cm}{
\parbox[t]{7.0cm}{
\includegraphics[width=7.0cm]{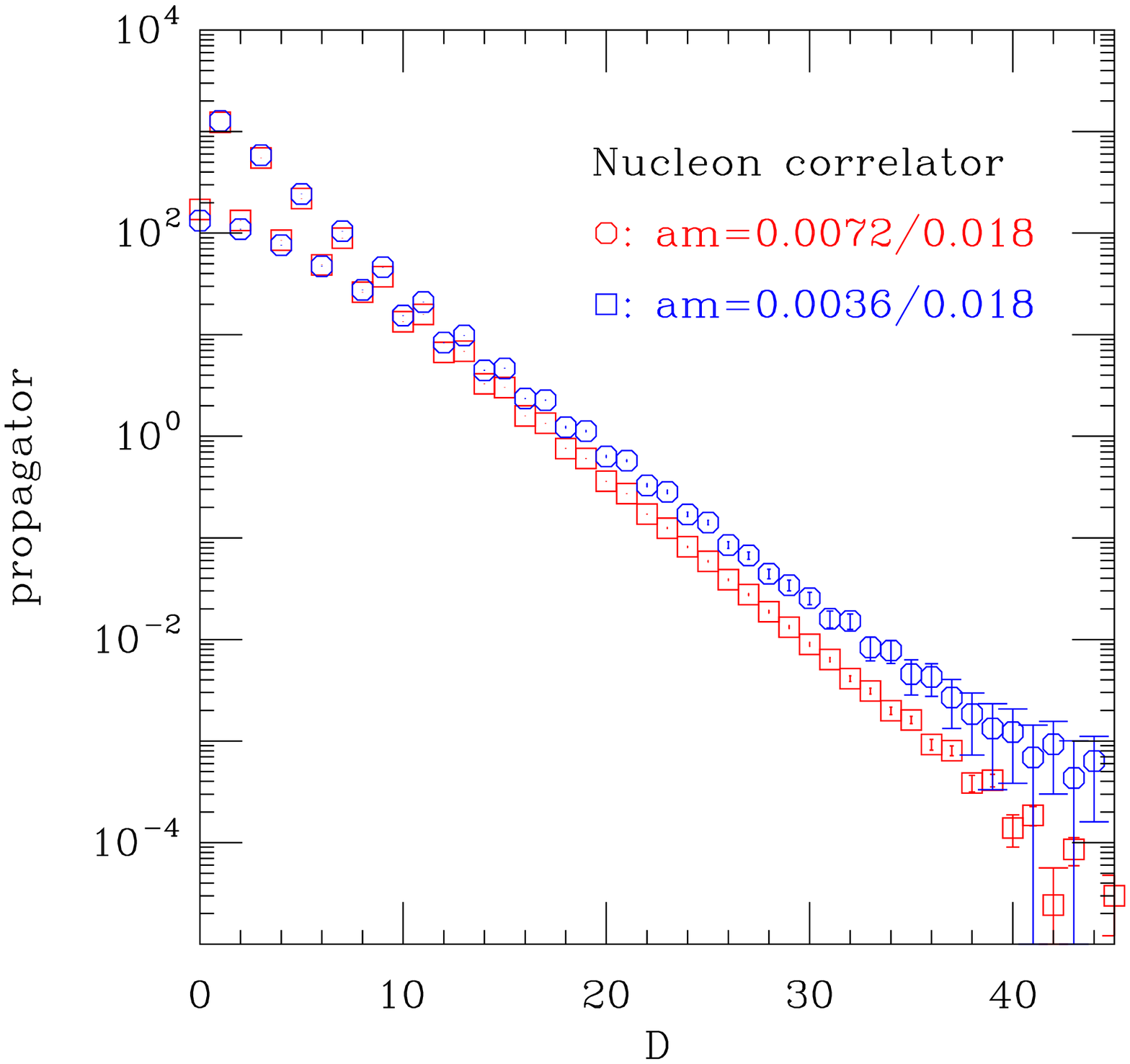}
Figure 3a:
Nucleon correlators at $a=0.06$ fm for light quark masses $0.4\,m_s$ and
$0.2\,m_s$.
} %end parbox
\hspace{1.0cm}
\parbox[t]{7.0cm}{
\includegraphics[width=7.0cm]{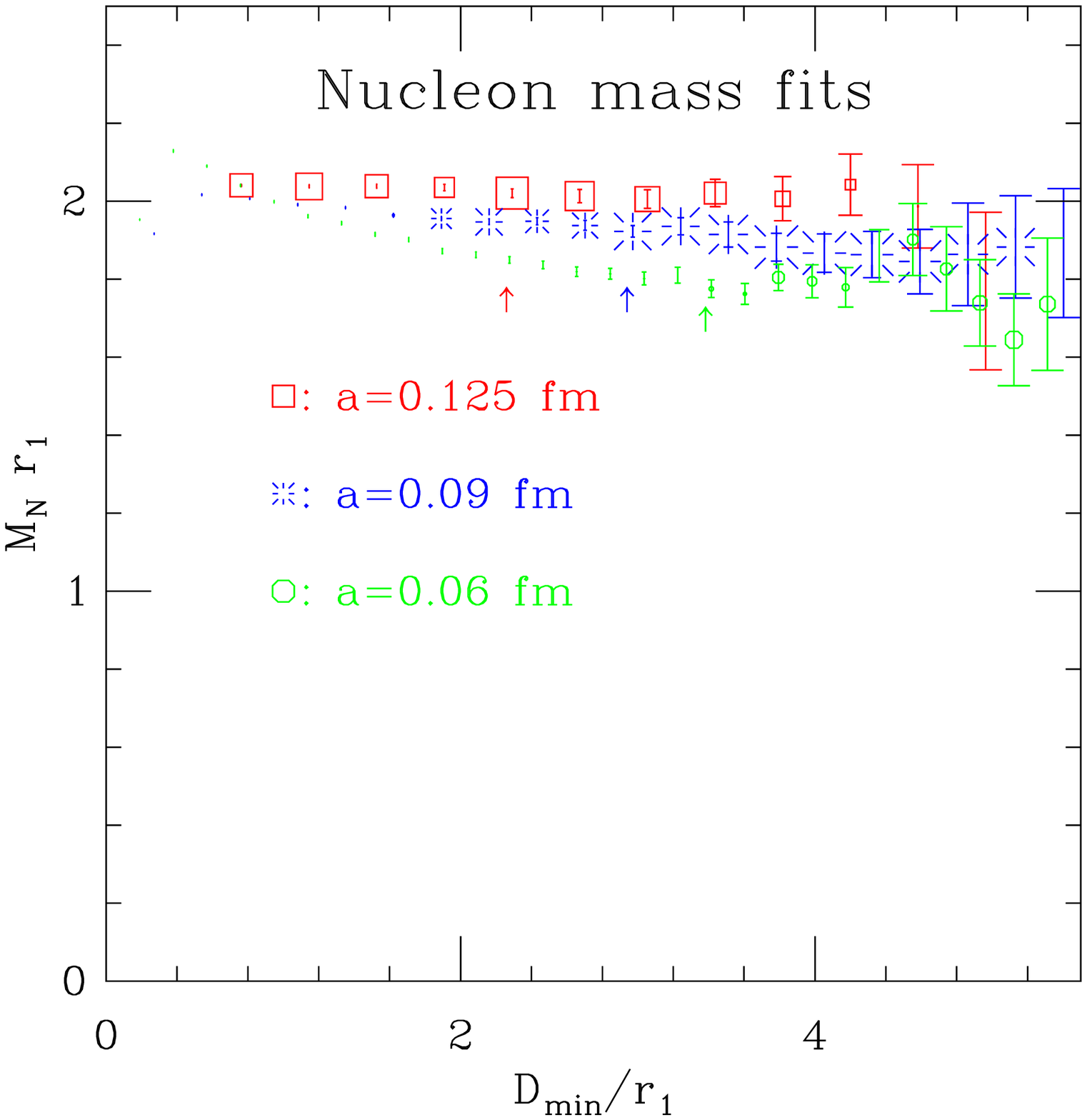}
Figure 3b:
Nucleon mass fits for $m_l = 0.2 m_s$.
}
}\end{center}

%\begin{center}\parbox[t]{15cm}{
%\parbox[t]{7.0cm}{
%\includegraphics[width=7.0cm]{mnuc_data.ps}
%Nucleon masses versus quark mass.
%}
%\hspace{1.0cm}
%\parbox[t]{7.0cm}{
%\includegraphics[width=7.0cm]{mnuc_data_zoom.ps}
%The same, expanded to show the small quark mass region.
%}
%}\end{center}

\begin{center}\parbox[ht]{15cm}{
\parbox[ht]{7.0cm}{
\includegraphics[width=7.0cm]{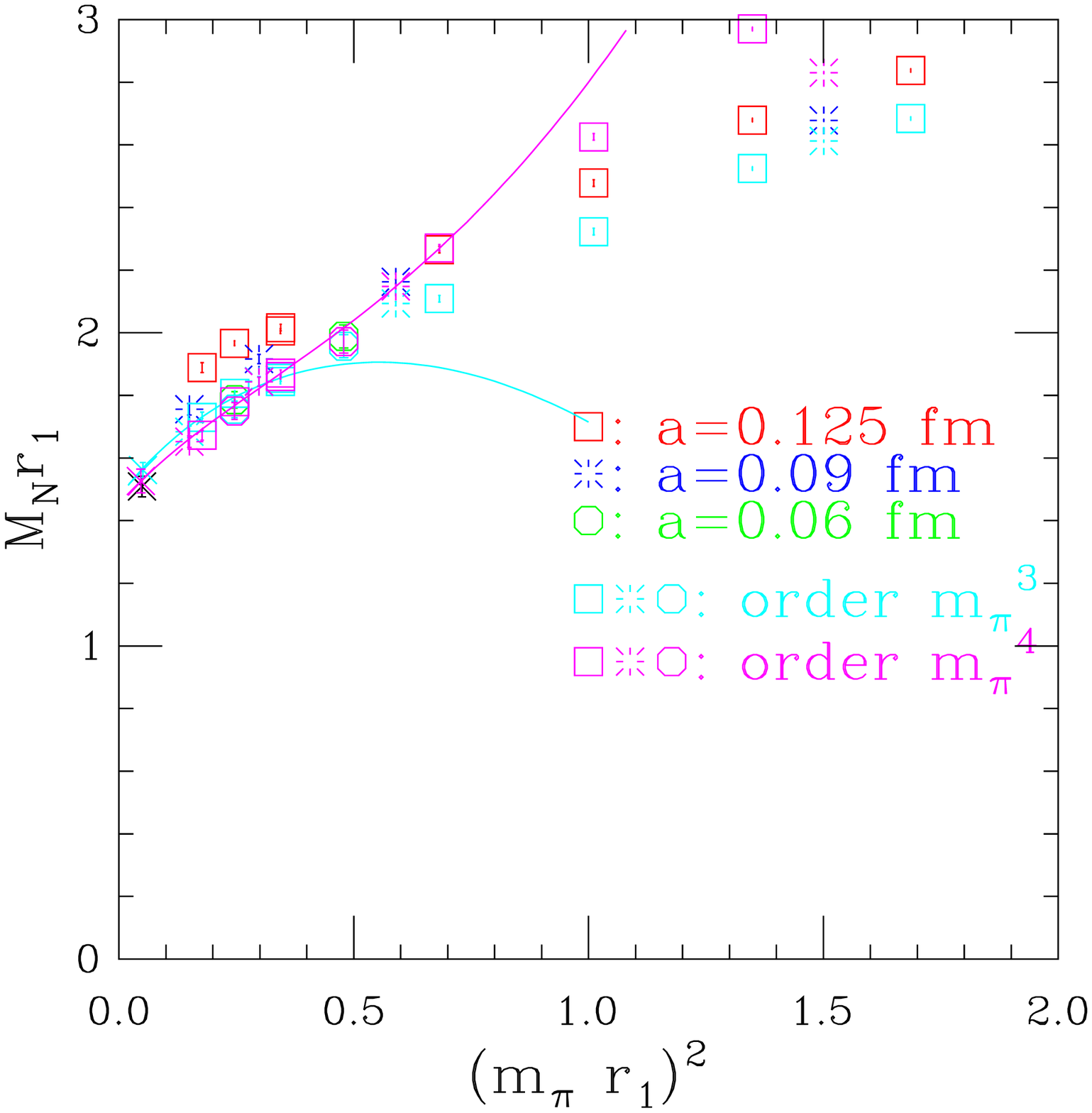}
Figure 4a:
Nucleon masses
(red blue and green symbols),
and fits to the nucleon mass (cyan and magenta).
}
\hspace{1.0cm}
\parbox[ht]{7.0cm}{
\includegraphics[width=7.0cm]{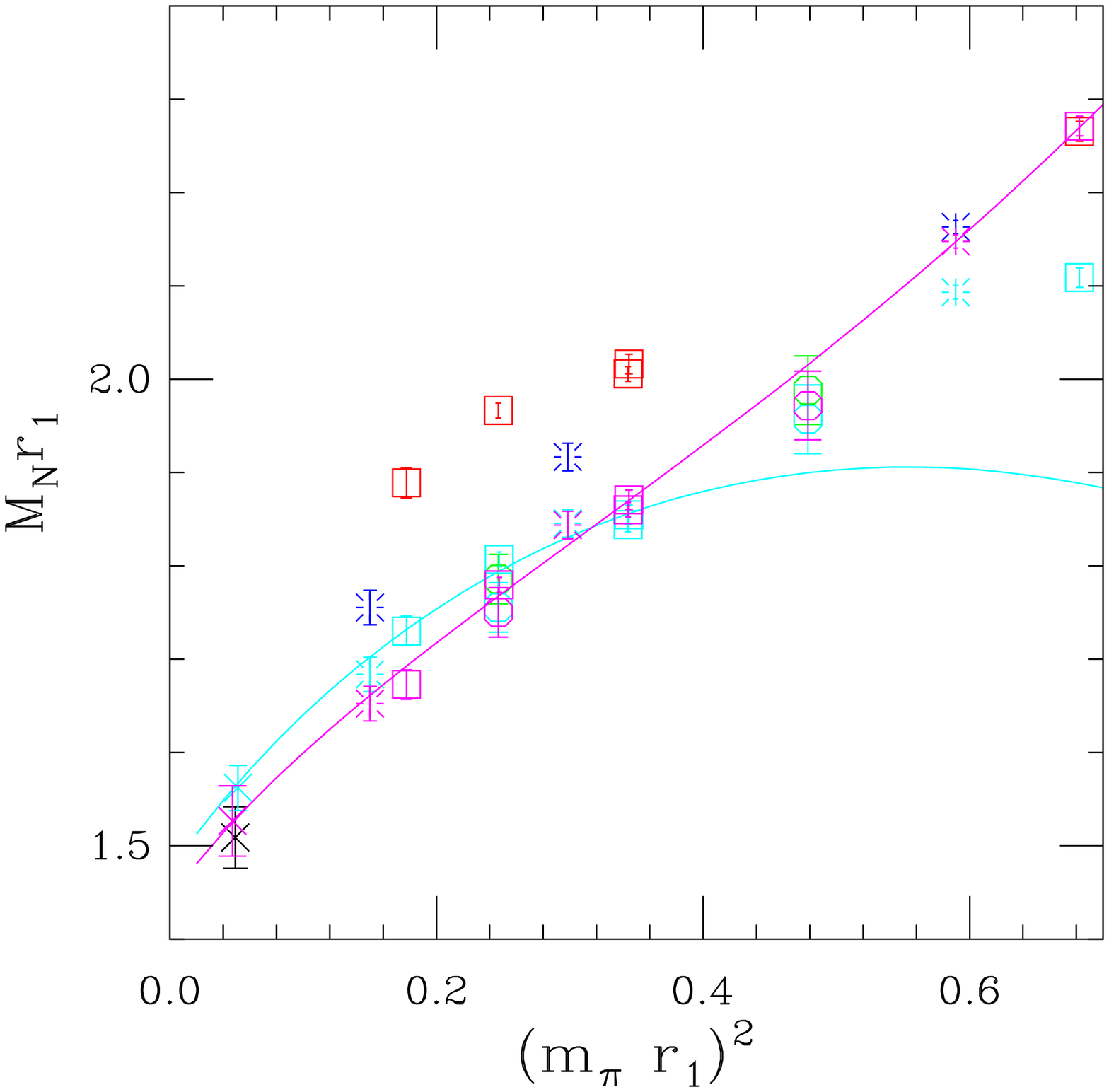}
Figure 4b:
The same, expanded to show the small quark mass region.
The points with $m_l=0.2\, m_s$ are from the fits indicated
with arrows in Fig.~3b.
}
}\end{center}

We show two forms for extrapolating the nucleon mass to the physical quark mass and
continuum limit.  The first (cyan) includes only the lowest order nonanalytic
correction, order $m_\pi^3$.

\BNE m_N r_1 = M_0 r_1 + A_1 (m_\pi r_1)^2 + B_1 a^2\alpha + r_1 \frac{-3
g_A^2}{32 \pi f_\pi^2} m_\pi^3\ENE

In these fits we just set $f_\pi$, $g_A$ and $M_\Delta-M_N$ to their physical values.
The cyan and magenta symbols in Figs.~4a and 4b are the data points with
the fitted lattice spacing corrections, $B_1 a^2\alpha$, subtracted, {\it i.e.} at $a=0$, and the cyan
and magenta lines are the fitting function at $a=0$.

The second form (magenta)\cite{MNUCCHIRAL} includes the $\Delta$ without the
assumption that $\Delta = M_\Delta-M_N >> f_\pi$.  When Taylor expanded in powers of $m_\pi$
this includes terms up to $m_\pi^4$, $m_\pi^4 \log(m_\pi)$.
\BNE
m_N r_1 = M_0 r_1 + A_1 (m_\pi r_1)^2  + A_2 (m_\pi r_1)^4 +
B_1 a^2\alpha + B_2 (m_\pi r_1)^2 a^2 \alpha 
 + r_1 \frac{6 g_A^2}{25\pi^2 f_\pi^2} H(m_\pi,M_\Delta-M_N,\Lambda)
 \ENE
where, if $m_\pi < \Delta = M_\Delta-M_N$
\BEA
&&H(m_\pi,\Delta,\Lambda) = \Delta^3 \log(2\Delta/m_\pi) +
    \Delta m_\pi^2 \LP \frac{3}{2}\log(m_\pi/\Lambda)-1 \RP \\
 &-& \LP \Delta^2-m_\pi^2\RP^{3/2} \log\LP \Delta/m_\pi +
\sqrt{\Delta^2/m_\pi^2 - 1} \RP
 - \Delta m_\pi^2 \LP \frac{3}{2}\log( 2\Delta/\Lambda ) -
 \frac{3}{4} \RP
\EEA
if $m_\pi>\Delta$
\BEA
&&H(m_\pi,\Delta,\Lambda) = \Delta^3 \log(2\Delta/m_\pi)
 + \Delta m_\pi^2 \LP \frac{3}{2}\log(m_\pi/\Lambda)-1 \RP \\
 &-& \LP m_\pi^2 - \Delta^2\RP^{3/2} \arccos( \Delta/m_\pi )
 - \Delta m_\pi^2 \LP \frac{3}{2}\log( 2\Delta/\Lambda ) -
 \frac{3}{4} \RP
\EEA

Systematic errors have not been properly analyzed yet, and the data
set is incomplete.  With these reservations, the order $m_\pi^4$ chiral
fit in Fig.~4 at the physical quark mass is
$M_N r_1 = 1.52 \pm .04 \pm {\rm syst.}$, or
$M_N(MeV) = 942 \pm 25 \pm {\rm syst.}$, where systematic errors
on $M_N r_1$ must be at least 0.045, the difference between the two
chiral fitting forms.  We are also experimenting with chiral extrapolations
using fitting forms from partially quenched staggered chiral perturbation
theory\cite{BAILEYTHESIS}.

In fitting the $\Omega^-$ mass the chiral extrapolation is simpler,
but we must deal with its dependence on the strange quark mass.
Our lattice simulations were run with an estimate of the strange quark
mass and only after running the simulation cam we determine what the
correct strange quark mass is.   Thus, we
compute correlators at two valence strange quark masses, and then
interpolate to the valence strange quark mass determined from pseudoscalar
mesons (basically $M_K$).
Figure~5a (like figure 3b for the nucleon) shows fits to the $\Omega^-$
correlators for runs at three different lattice spacings with $m_l \approx 0.2 m_s$,
again with arrows indicating the fit we chose.
For the chiral and continuum extrapolation we use
a linear extrapolation in $am_l$ and $a^2\alpha$, which fits well
with $\chi^2/dof = 6.0/7$.
\BNE M_\Omega r_1 = M_0 r_1 + A (m_\pi r_1)^2 + B a^2 \alpha \ENE

Figure~5b shows the $\Omega^-$ masses interpolated to the strange quark
mass determined from pseudoscalar mesons as a function of light quark mass.
It also shows the fit form above for $a\approx 0.125$, $0.09$ and
$0.06$ fm, and in the continuum limit, {\it i.e.} with
$B$ set to zero. 
The black symbol is the experimental value for $m_\Omega r_1$, where the error is
from the uncertainty in $r_1$.

Errors on the extrapolated lattice result are statistical (0.7\%),
determination of $m_s$ from pseudoscalars (0.6\%),
and an estimate of neglected NNLO terms in chiral extrapolation ($\null_{0.05}^{0}$).
Other small errors come from using the wrong strange sea quark mass, and the choice
of procedure for fixing strange quark mass.
Using data available at conference time, the continuum and chiral extrapolated
value is $M_\Omega * r_1 = 2.679 {{+.025}\atop{-.056}}$.
Converting to physical units using $r_1 = 0.318(7)$ and adding errors, this is
$M_\Omega(MeV) = 1660 {{+40}\atop{-50}}$, where the experimental value is 1672 MeV.

This work is supported by the US Department of Energy and National Science Foundation.
Computations were done at the NSF Teragrid, NERSC, USQCD centers and computer centers
at the Universities of Arizona, Indiana, Utah and California at Santa Barbara.

\begin{center}\parbox[ht]{15cm}{
\parbox[ht]{7.0cm}{
\includegraphics[width=7.0cm]{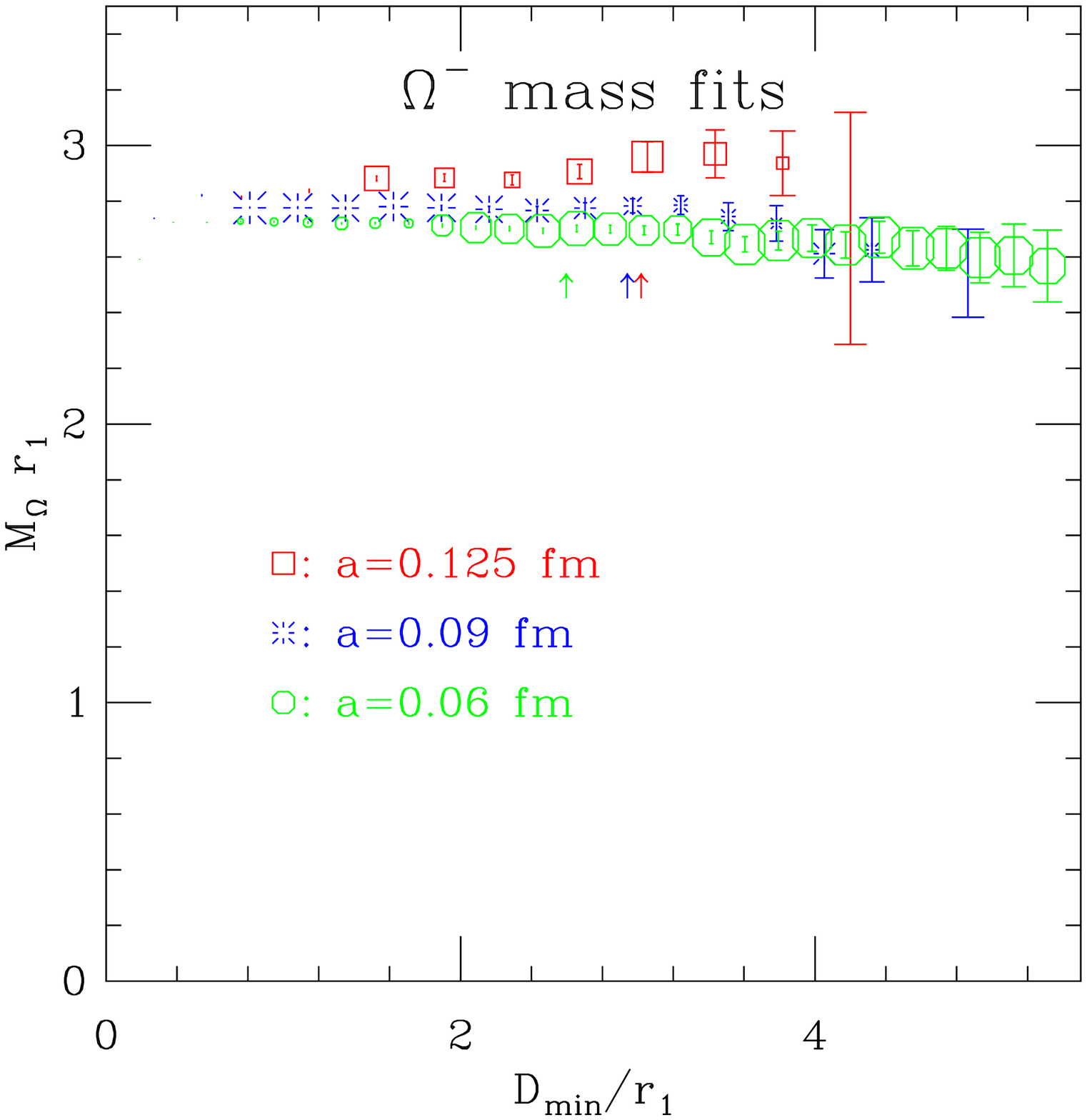}
Figure 5a:
$\Omega^-$ mass fits for $m_l = 0.2 m_s$.  In these plots
the strange quark mass has
not been tuned to the correct $m_s$; it is one of the masses
at which correlators were computed.
}
\hspace{1.0cm}
\parbox[ht]{7.0cm}{
\includegraphics[width=7.0cm]{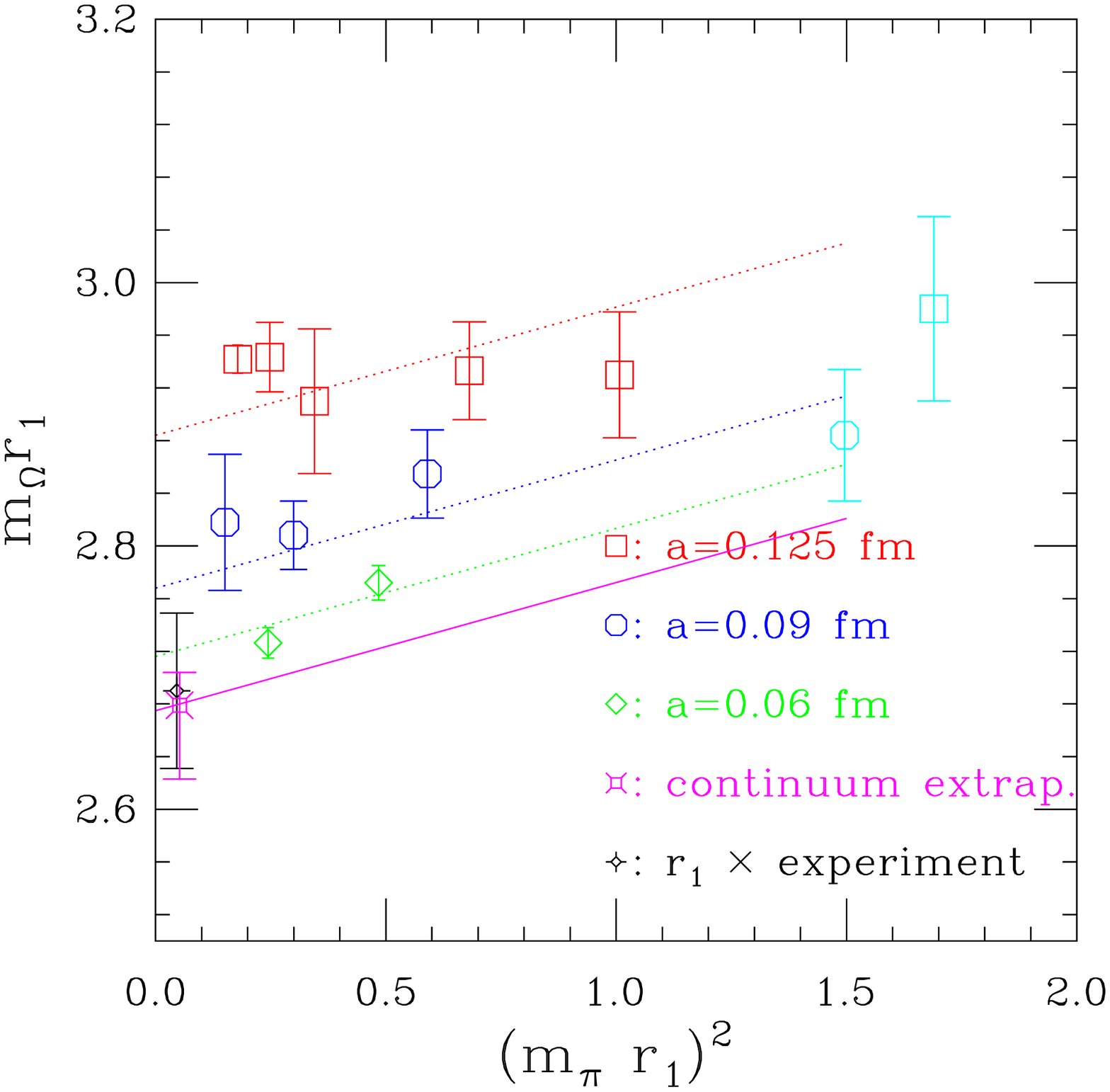}
Figure 5b:
$\Omega^-$ masses and fitting (see text).
The error on the experimental point is from the error
in $r_1$.
}
}\end{center}

\end{document}